\documentclass{zArxiv-ws_mpla}
\usepackage[super]{cite}
\usepackage{graphicx}
\usepackage{enumerate}
\usepackage{mathtools}

\DeclarePairedDelimiter\abs{\lvert}{\rvert}%
\begin{document}

\markboth{B.D.~Normann and I.H.~Brevik}
{Can the Hubble tension be resolved by bulk viscosity?}

\catchline{}{}{}{}{}

\title{Can the Hubble tension be resolved by bulk viscosity?}

\author{\footnotesize Ben David Normann\footnote{corresponding author}}

\address{Department of ICT and Natural Sciences, Norwegian University of Science and Technology, Postboks 1517 NO-6025\\
6025 Ålesund, Norway\\
ben.d.normann@ntnu.no}

\author{Iver Håkon Brevik}

\address{Department of Energy and Process Engineering, Norwegian University of Science and Technology, Postboks 8900, Torgarden\\
7491 Trondheim, Norway\\
iver.h.brevik@ntnu.no}

\maketitle

\pub{Received (Day Month Year)}{Revised (Day Month Year)}

\begin{abstract}
We show that the cosmic bulk viscosity estimated in our previous works is sufficient to bridge the $H_0$ value inferred from observations of the early universe with the value inferred from the local (late) universe.

\end{abstract}

\ccode{PACS Nos.: include PACS Nos.}

\section{Introduction}
Bulk viscosity has been a popular research field of theoretical cosmology since its early relativistic formulation by  Israel, Stewart,  Weinberg and other authors.  In the early seventies Suszycki, Klimek and Heller proposed for the first time a bulk-viscous means of avoiding the initial singularity~\cite{heller73}\footnote{Also note the reference therein to Klimek's prior theoretical work\cite{Klimek71} (in Russian).}. Since these early developments the literature has been abundant with contributions. For instance, Fabris \cite{fabris06} et al. explored in 2006 the possibility of the late acceleration being an artefact of a cosmological viscosity. A host of subsequent studies have been performed. We suffice it here to mention the recent work by Hu and Hu\cite{hu20}, who report that a cosmological viscosity in combination with a cosmological constant is superior in describing observational data compared to the $\Lambda$CDM model.

In the most recent literature, the $H_0$-tension has also been sought alleviated through bulk viscosity. The $H_0$-tension arises as a result of discrepancies in the inferred value of $H_0$ from different types of measurements. On the one hand one finds the results of local measurements such as Cepheid-calibrated supernovae~\cite{riess19,reid19}, strong-lensing time-delays~\cite{wong19}
and others. These model-independent local measurements yield a consistent value of about $H_0\,\sim\,$73$~{\rm km\, s}^{-1}{\rm Mpc}^{-1}$. On the other hand one finds values inferred within the $\Lambda$CDM-paradigm from the cosmic microwave background radiation (CMB)~\cite{planck20}, yielding a value for $H_0\,\sim\,$ 69$~{\rm km\,s}^{-1}{\rm Mpc}^{-1}$. The  recent review by Valentino et al.  ~\cite{divalentino21_realm}  gives  a thorough explanation of the different kinds of data and the proposed solutions. See also the recent paper by Macorra, Almarez and Garrido~\cite{delamacorra21}, where a a solution is sought through adding extra relativistic energy to the early universe. Yet another, very recent strategy to resolving the tension, is to make use of machine learning to narrow down the range of parameter values in a two-component fluid system~\cite{elizalde21}. Nojiri and others~\cite{nojiri21} recently suggested dealing with the Hubble tension through a modified equation-of-state approach. To this end, it is very interesting to note Velten, Costa and Zimdahls approach, which we were just recently made aware of~\cite{velten21}. The bulk-viscous contribution to the energy density generated by a two-component fluid (radiation and dark matter) is estimated at early times ($z\,\sim\,3400$).

Assuming that the discrepancy is not due to systematic error somewhere along any of the pipelines, the $H_0$-tension is  one of the great mysteries of modern cosmology. As is customary, we interpret redshift to be due to cosmic expansion. Consequently, the high-redshift observations are from early times ($z \approx 1100$) whereas the low-redshift observations are from more recent epochs  ($z<2.3$). Hence we shall in the following refer to the value inferred from the first category of experiments as the \textit{high-redshift} (high-$z$) value, whereas the value inferred by the latter category is referred to as the \textit{low-redshift} (low-$z$) value. In agreement with the mentioned review~\cite[Fig.~2]{divalentino21_realm} we adopt
\begin{align}
   &H_{0\,\rm high-z}=69.3\,{\rm km}\,{\rm s}^{-1}{\rm Mpc}^{-1},\\
    &H_{0\,\rm low-z}=73.2\,{\rm km}\,{\rm s}^{-1}{\rm Mpc}^{-1}.
\end{align}
Furthermore, we define for convenience  the factor $r$ representing the ratio of these best-fit values,
\begin{align}
    r=H_{0\,\rm low-z}/H_{0\,\rm high-z}=1.056.\label{r}
\end{align}
The solution to the   $H_0$-tension  problem is as of yet not known, and many different approaches have been taken to solve it. For instance, a recent paper seeks to ameliorate the tension by studying the effects of two flows within the $\Lambda$CDM paradign.  Another, very recent paper confronts a viscous Chaplygin-gas model with an array of observations and suggests that it may not only describe the Universe's accelerated expansion, but also ameliorate the tension~\cite{hernandezAlmada21}.

In addition to  the $H_0$-tension, the large-scale structure extrapolated from the CMB, namely the  $\sigma_8$-value, is in conflict with that of other LSS observations. Interestingly, Anand et al has shown that this tension may be remedied through viscous effects in the cosmic fluid~\cite{anand17}.  Actually, a small viscosity in the cosmic fluid is argued to be preferable.  In a subsequent paper, such viscosity is also noted  to strongly constrain neutrino masses~\cite{anand18}.

It has also been shown that the $\sigma_8$ discrepancies between LSS and CMB  may vanish when considering the energy-momentum tensor for an imperfect fluid~\cite{mohanty18}. The authors demonstrate how the presence of viscosities in the cold, dark fluid on large scales ameliorate the problem `more elegantly than other solutions', to use the words of the authors.  Interestingly, the authors are also able to obtain an estimate of the neutrino mass in their viscous cosmology. Alongside the same line one may observe the work by Mishra~\cite{mishra20a}, in which the self-interaction of dark matter (SIDM), with corresponding viscosities, is calculated from kinetic theory. Although small, the SIDM viscosity is reported to contribute significantly to the cosmic dissipation budget at low redshift. The paper also describes how the visible photon production from the dark matter (DM) fluid depends on the amount of DM viscosity~\cite{mishra20b}, which therefore may be inferred from measurements. Indeed, this is reported to have the capacity to explain the EDGES\setcounter{footnote}{0}\footnote{EDGES is an acronym for \textit{Experiment to detect the Global EoR Signature}. EoR stands for \textit{Epoch of Reionization.}} anomalies. See also~\cite{bhatt19,mishra20c,atreya18,atreya19,natwariya20} for more information about  this line of research.

Sasidharan et al.~\cite{sasidharan18} report that a certain bulk-viscous matter model is marginally preferred over the concordance model. The functional form of the viscosity-parameters used therein is $\zeta=\zeta_0+\zeta_1 H$, where the parameters are  dimensional constants and $H$ is the Hubble parameter. Since $\rho\sim H$, this is in good agreement with the functional form $\zeta\sim\rho^{1/2}$ which also we give preference to~\cite{Normann16,normann17,brevik20}. This should not come as a surprise, since this is the functional form that naturally would result if the viscosity is to remain within a first-order deviation from the barotropic pressure of a perfect fluid for all times~\cite{shogin15,jerinMohanN18}. Yet another group, Yang et al., finds that a power-law bulk viscosity is always preferred over $\Lambda$CDM, but the latter is preferred in a Bayesian analysis~\cite{yang19}.

Even more recently, Elizalde and others~\cite{elizalde20} study a two-component inhomogeneous model~\footnote{Inhomogeneous in this context refers to the type of equation-of-state, and not the symmetries of space-time}. Again the conclusion is that bulk viscosity may remedy the tension.

In the present  note, we   report what conclusions can be drawn  on the basis of our  previous works on bulk-viscous cosmology~\cite{Normann16,normann17,brevik20}, where we estimated the magnitude of the present-day bulk viscosity in the cosmic fluid. What we in particular would like to focus on, is whether or not such a viscosity is sufficiently large to to ameliorate the $H_0$-tension.  Our  picture is  the following: the low-$z$ value of $H_0$ is identified as the ``true" Hubble parameter, as this case lies relatively close to us in time. The  viscosity has not yet had sufficient time to work. The high-$z$ case is  however different, as there is ample time for the viscosity to become noticeable.   We wish to test if the mentioned known value of the bulk viscosity is sufficient to provide a natural explanation of the increase of  $H_0$,  from the high-$z$ value to the low-$z$ value. And actually the formalism strongly indicates that it may be so.   An important point in our analysis is that, once the value of the viscosity is chosen, there are no other adjustable parameters.

\section{Calculations}
In our previous work~\cite{Normann16} we found that the dimensionless Hubble parameter $E^2=H^2/H_0^2$ is given by
\begin{align}
    E^2(z)=\Omega(z)(1+u(z,B,\lambda)).
\end{align}
In the above, the normalized matter density is given by
\begin{align}
    \Omega(z)=\sum_i\Omega_{i0}(1+z)^{3(\omega_i+1)}\label{omega},
\end{align}
where and $\Omega_{i0}=\rho_{i0}/\rho_{\rm c}$ is the present-day value and $\rho_{\rm c}=3H_0^2/8\pi\,G$ is the critical density at present. Since $H_0$ is influenced by any viscosity added to the cosmic fluid, so is the critical density, in general.  
The viscosity-factor $u(z,B,\lambda)$ is given by
\begin{equation}
\label{C_Sol_u_final}
{
u(z,B,\lambda)=
\begin{cases}
\displaystyle\left[ 1-\left(1-2\lambda\right)\frac{B}{H_0}\int_0^z{\frac{1}{(1+z)\sqrt{\Omega}^{1-2\lambda}}}dz\right]^{\frac{2}{1-2\lambda}}-1
& \phantom{00}\text{for}\text{~}\lambda\neq \frac{1}{2},\\
\displaystyle (1+z)^{-\frac{2B}{H_0}}-1
& \phantom{00}\text{for}\text{~}\lambda=\frac{1}{2},\\
\end{cases}
}
\end{equation}
(there is a misprint in our original source for the case $\lambda\,\neq\,1/2$).  The following definition is useful,  
\begin{align}
    B=12\pi\,{\rm G}\zeta_0.\label{usefuldef}
\end{align}
In \cite{Normann16} we found, based upon the assumption $ k=0$, that the best-fit value of the present-day viscosity for $\zeta\propto\sqrt{\rho}$ is $B\,=\,0.76 {\rm km}\,{\rm s}^{-1}{\rm Mpc}^{-1}$. Using~\eqref{usefuldef} we find that this corresponds in more familiar units to $\zeta_0$ slightly short of $10^6\,{\rm Pa\,s}$.
This estimate seems to agree well enough with other studies~\cite{brevik15,wang14,velten12,sasidharan16,brevik16}.

As argued earlier, we give preference to $\zeta\,\sim\,\sqrt{\rho}$. In the following, we therefore assume such a functional form for the viscosity $\zeta$. The development over redshift for the viscous and non-viscous Hubble parameters is thus
\begin{align}\label{Hb}
&H_{\rm B}^2(z)=H_{0\rm B}^2\Omega_{\rm B}(z)(1+z)^{-\frac{2B}{H_0}},\\
&H^2(z)=H_{0}^2\Omega_{\rm}(z)\label{H},
\end{align}
respectively. 
The present-day Hubble parameter takes different values in the two different scenarios.

\section{Can bulk viscosity account for the Hubble tension?}
The question we intend to address in this  note is, as mentioned,  whether or not the model described in the previous section is capable of accounting for the observed tension in the present-day Hubble parameter $H_0$.

We then need to investigate if the viscous modification can translate the high-z value into  the low-z value inferred from the second category of experiments. In the following, we let $\tilde{z}$ refer to the value of the redshift parameter $z$ at the time of recombination. Consequently adopt the following two requirements.
\begin{enumerate}[(1)]
\item We require that the Hubble parameter at recombination (CMB) is the same for the viscous and non-viscous models.
\begin{align}
&H_{\rm B}(\tilde{z})=H(\tilde{z})\label{req1}.
\end{align}
This assumption is made in order to calculate the viscous contribution from recombination and onward.
\item  As mentioned, we  require that
\begin{align}
&H_{0\rm B}=H_{0\,\rm low-z}\quad\quad\quad\quad\textrm{and}\quad\quad\quad\quad H_0=H_{0\,\rm high-z}\label{req2}.
\end{align}
This requirement encodes the assumption we implicitly want to make; that the low-z (local Universe) measurements indicate the correct value of $H_0$, whereas the high-z values are wrong due to the omitting of viscous contributions in standard cosmology. Thus we are equipped to quantify the viscosity needed in order to bridge the two values.
\end{enumerate}
Note that $H_{0B}$  and  $H_0$  are theoretical quantities, while $H_{0\rm low-z}$ and $H_{0\rm high-z}$  are experimental ones.
Starting from Eq.~\eqref{req1}, using Equations~\eqref{Hb}-\eqref{H} and the requirement~\eqref{req2}, one finds the expression
\begin{align}
    B=-\frac{H_{\rm 0B}}{2}\frac{\ln{A(\tilde{z})}}{\ln{(1+\tilde{z}})}\label{B}
\end{align}
for the astronomical viscosity-parameter $B$. Here 
\begin{align}
    A(z)=\frac{\rho_{\rm hom}(z)}{(\rho_{\rm B})_{\rm hom}(z)}.
\end{align}
Here $\rho_{\rm hom}(z)$ and $(\rho_{\rm B})_{\rm hom}(z)$ are the matter densities of the non-viscous and viscous universe, respectively. The sub-script  ``$\rm hom$" indicates that this is the \textit{homogeneous} solution to the energy-conservation equation. Physically, this corresponds to omitting the viscosity\setcounter{footnote}{0}\footnote{But since the initial conditions are met differently when including viscosity, the homogeneous part of $\rho_{\rm B}(z)$ will nevertheless yield different values over $z$ compared to the non-viscous scenario; $\rho(z)$. Thus $A(z)$ is generally not 1.}. The accurate expressions for the $\rho$s are not of any importance here, but may be inferred from Equation~\eqref{omega}. Besides, they are given in our previous work. 
Next, we are confronted  with two possibilities.
\begin{enumerate}[i]
    \item \textbf{Non-viscous scenario} (trivial): Requiring \begin{align}
    (\rho_{\rm B})_{\rm hom}(\tilde{z})=\rho_{\rm hom}(\tilde{z})
    \end{align}
    This would give $A(\tilde{z})=1$ and hence $B=0$. Thus it corresponds to the trivial solution, and would not be capable of explaining any difference in the two values of $H_0$.
    \item \textbf{Viscous scenario}:  Require instead that the normalized matter densities of the viscous and non-viscous models  coincide at recombination,
    \begin{align}
    (\Omega_{\rm B})(\tilde{z})=\Omega(\tilde{z}).
    \end{align}
    This  results in $A(\tilde{z})=H_0^2/H_{0B}^2=r^{-2}$, where $r$ is defined in~\eqref{r} and ~\eqref{req2} is adopted.
\end{enumerate}
With option (ii) above, Eq.~\eqref{B} becomes
\begin{align}
    B=H_{\rm 0B}\frac{\ln{\abs{r}}}{\ln{(1+\tilde{z}})},
\end{align}
Inserting now $r=1.056$ and $\tilde{z}=1100$ we find 
\begin{align}
    B\,=0.572\,{\rm km}\,{\rm s}^{-1}{\rm Mpc}^{-1}.
\end{align}
What we are looking at here, is an integrated viscosity contribution from $z\sim 1100$ til present, and the big value of $z$ is thus important, as a smaller history (lesser value of $\tilde{z}$) would lead to a bigger numerical value for $B$. One may nevertheless note that the result is not overly sensitive to variations in $\tilde{z}$, the mathematical reason for this being that this number is already very large.

Note, however, that the result is much more delicately dependent on $r$, as this value is close to 1. Hence $\ln{r}$ hovers above 0. Pushing $r$ a bit further towards 1 will cause a relatively big change in $\ln{r}$ and thus in $B$.

Whereas the best-fit viscosity calculated in our previous work was based on a particular model for the matter content, the results of this paper are quite model independent,  as we merely require the viscosity to be associated with the fluid as a whole and not only with one of its components. Hence, a perfect agreement would seem suspicious. However, we nevertheless find it interesting that the agreement with our previous work (where we as mentioned found $B\,=0.755\,{\rm km}\,{\rm s}^{-1}{\rm Mpc}^{-1}$ as a best-fit value) is of the same order of magnitude. And most importantly: The viscosity peviously found is sufficient. 

As for the microscopic causes behind the viscosity, investigations along the same lines as that of Velten, Costa and Zimdahl~\cite{velten21} seem highly relevant. It is also instructive to note the recent study of the Pantheom sample by Dr. Dainotti and others~\cite{dainotti21}, in which they find a trend of decreasing Hubble-parameter value as a function of redshift. We suggest that bulk viscosity may be the phenomenological explanation of this trend.    

\section{Conclusions}
In this paper we have shown that bulk-viscous corrections to the Hubble flow, as calculated through the formalism derived in previous work~\cite{Normann16}, is of sufficient magnitude to ameliorate the observed tension in the high-redshift versus low-redshift values of $H_0$. Recall that, once the value for the bulk viscosity is inserted, the formalism contains no further adjustable parameters.  Our value for the bulk viscosity  is in agreement  also with other recent  statements in  the literature. 

\newpage
\bibliographystyle{ws-mpla}
\bibliography{Viscrefs}

\end{document}